\documentclass[fleqn,pra]{revtex4}

\usepackage{amsmath,graphicx}
\usepackage{dcolumn}
\usepackage{verbatim}

\begin{document}

\title{Relativistic corrections of $m\alpha^6(m/M)$ order to the
hyperfine structure of the $\mbox{H}^+_2$ molecular ion}

\author{Vladimir I. Korobov}
\affiliation{Bogolyubov Laboratory of Theoretical Physics, Joint Institute
for Nuclear Research, Dubna 141980, Russia}
\author{L.~Hilico}
\author{J.-Ph.~Karr}
\affiliation{Laboratoire Kastler Brossel, UEVE, CNRS, UPMC, ENS\\
D\'epartement de Physique et Mod\'elisation, Universit\'e d'Evry Val d'Essonne, Boulevard
F.~Mitterrand, 91025 Evry Cedex, France}

\begin{abstract}
The $m\alpha^6(m/M)$ order corrections to the hyperfine splitting in the
$\mbox{H}_2^+$ ion are calculated. That allows to reduce uncertainty in
the frequency intervals between hyperfine sublevels of a given
rovibrational state to about 10 ppm. Results are in good agreement with
the high precision experiment carried out by Jefferts in 1969.
\end{abstract}

\maketitle

\section{Introduction}

In our previous work \cite{HFS06} we have calculated the hyperfine structure of the hydrogen molecular ion $\mbox{H}_2^+$ within the Breit-Pauli approximation taking account of the anomalous magnetic moment of an electron. This approximation includes the contributions of order $m\alpha^4(m/M)$ and $m\alpha^5(m/M)$ and thus the relative uncertainty in determination of the hyperfine structure intervals is of about $5\times10^{-5}$. For the first time that has allowed to confirm the Jefferts measurements \cite{Jeff69} to the level of experimental accuracy of 1.5 kHz for transitions within the same multiplet $F$ ($F$ is the total spin of a state in ion). For the spin-flip transitions
$(F=3/2)\to(F=1/2)$ a discrepancy of about 80 kHz still remains.

The main goal of the present work is to consider higher order corrections
to the hyperfine splitting of $\mbox{H}_2^+$ to reduce the discrepancy
with the Jefferts experiment for spin-flip lines down to a few ppm. To
that end we will calculate the QED contributions of order $\alpha^2E_F$
and partially of order $\alpha^3E_F$ along with the proton finite size
corrections such as Zemach and pure recoil contributions, which are
essential at this level of accuracy.

The effective Hamiltonian of the spin interaction for the $\mbox{H}_2^+$
ion is (we use notation of \cite{HFS06}):
\begin{equation}\label{effH_H2}
\hspace{-5mm}
\begin{array}{@{}l}
\displaystyle H_{\rm eff} =
    b_F(\mathbf{I}\cdot\mathbf{s}_e)
   +c_e(\mathbf{L}\cdot\mathbf{s}_e)
   +c_I(\mathbf{L}\cdot\mathbf{I})
   +\frac{d_1}{(2L\!-\!1)(2L\!+\!3)}
       \biggl\{
          \frac{2}{3}\mathbf{L}^2(\mathbf{I}\cdot\mathbf{s}_e)
          -[(\mathbf{L}\cdot\mathbf{I})(\mathbf{L}\cdot\mathbf{s}_e)
           \!+\!(\mathbf{L}\cdot\mathbf{s}_e)(\mathbf{L}\cdot\mathbf{I})]
       \biggr\}
\\[3mm]\displaystyle\hspace{60mm}
   +\frac{d_2}{(2L\!-\!1)(2L\!+\!3)}
       \left[
          \frac{1}{3}\mathbf{L}^2\mathbf{I}^2
          -\frac{1}{2}(\mathbf{L}\cdot\mathbf{I})
          -(\mathbf{L}\cdot\mathbf{I})^2
       \right],
\end{array}\hspace{-10mm}
\end{equation}
here $\mathbf{I}$ is the total nuclear spin, $\mathbf{L}$ is the total
orbital momentum. The assumed coupling scheme of angular momenta is:
$\mathbf{F}=\mathbf{I}+\mathbf{s}_e$, $\mathbf{J}=\mathbf{L}+\mathbf{F}$.

The major coupling is the spin-spin electron-proton interaction (first
term in (\ref{effH_H2})) which determines the principal splitting between
$F=1/2$ and $F=3/2$ states. So, the main contribution to the theoretical
uncertainty on the spin-flip transition frequencies is uncertainty in the
spin-spin interaction coefficient $b_F$, and our aim is to calculate an
improved value for $b_F$.

Here it is very useful to make a comparison with the HFS studies of the
hydrogen atom ground state. Indeed, the analytical form of many
contributions to the hyperfine splitting of $\mbox{H}_2^+$ can be obtained
from these results. Moreover, we will use the known results on the
hydrogen atom as a guide and a check of our analytical derivations.

The hyperfine splitting for the ground state of a hydrogenlike atom may be obtained with high accuracy already from the nonrelativistic quantum mechanics (see for example \cite{BS}),
\begin{equation}
E_F =
   \frac{16}{3}\alpha^2cR_\infty\mu_p\frac{m_e}{M_p}
   \left[1+\frac{m_e}{M_p}\right]^{-3},
\end{equation}
here $\mu_p$ is the magnetic moment of a proton in nuclear magnetons, $m_e$ and $M_p$ are the electron and proton masses, respectively. Quantum electrodynamics corrections without recoil terms have been known for some time \cite{SapYen,Kin96} and may be expressed as:
\begin{equation} \label{qedcorr}
\Delta E_{\mathrm{hfs, QED}} =
   E_F\left[
      1+a_e+\frac{3}{2}(Z\alpha)^2
      +\left(\ln{2}-\frac{5}{2}\right)\alpha(Z\alpha)
      -\frac{8}{3\pi}\alpha(Z\alpha)^2\ln^2(Z\alpha)
      +\dots
   \right],
\end{equation}
where $a_e$ is the electron anomalous magnetic moment. We keep $Z$, the
nuclear charge number, in all expressions in order to make clear the
origins of different corrections.

Beyond pure QED corrections there are also proton structure effects (see \cite{SapYen,Shabaev,Carlson} for a detailed discussion). The leading one is the Zemach correction \cite{Zemach} ($(Z\alpha)(m/\Lambda)E_F$) that along with the radiative corrections to the nuclear structure contribution \cite{Karsh97} reads
\begin{equation}\label{eq:Zemach_at}
\Delta E_Z =
   -2\frac{m_eM_p}{M_p+m_e}\left(Z\alpha\right) R_Z(1+\delta_Z^{\rm rad})E_F,
\end{equation}
Here $R_Z$ is the Zemach radius, a mean radius associated with the
proton's charge-current distribution,
\[
R_Z=\frac{1}{\pi^2}\int \frac{d^3q}{q^4}
   \left[1-\frac{G_E(-q^2)G_M(-q^2)}{\mu_p}\right],
\]
where $G_E$ and $G_M$ are the electric and magnetic form factors of a
proton. We take $R_Z=1.045(16)$ fm \cite{Shabaev}. Radiative corrections
$\delta_Z^{\rm rad}$ to the Zemach contribution were obtained in
\cite{Karsh97} and $\delta_Z^{\rm rad}=0.0153$. The parameter $\Lambda$
determines the energy scale that corresponds to the mean radius of the
proton, and $\Lambda \approx 0.8 M_p$~\cite{Karsh97}. Next are the pure
recoil proton structure corrections of orders $(Z\alpha)^k(m/M)E_F$
($k=1,2$) \cite{BodYen88}
\begin{equation}\label{eq:recoil}
\Delta_R^p \approx (5.84\pm0.01)\!\times\!10^{-6}\>E_F.
\end{equation}
The last remaining effect of order $(Z\alpha)(m/M)E_F$, which has to be included, is the proton polarizability \cite{Fau02}
\begin{equation} \label{pol}
\Delta_{\rm pol} \approx (1.4\pm0.6)\!\times\!10^{-6}\>E_F.
\end{equation}

A summary of various contributions to the HFS of the hydrogen ground state
is given in Table \ref{tab:HFS_H}. Up to now, in our previous studies for
the $\mbox{H}_2^+$ ion \cite{HFS06}, only the contributions from the first
two lines have been included into consideration. In the present work we
intend to extend our research to higher order QED corrections (up to
$E_F\,\alpha(Z\alpha)^2\ln^2(Z\alpha)$ term) as well as the proton
structure effects.

\begin{table}[t]\label{tab:HFS_H}
\begin{center}
\begin{tabular}{c@{\hspace{10mm}}d}
\hline\hline
term & \multicolumn{1}{c}{(kHz)} \\
\hline
\vrule width 0pt height 10pt depth 3pt
$E_F$ & 1418\,840.09 \\
$a_e E_F$ & 1\,645.361 \\
$\Delta E_{(Z\alpha)^2}$ & 113.333 \\
$\Delta E_{\alpha(Z\alpha)}$ & -136.517 \\
$\Delta E_{\alpha(Z\alpha)^2\ln^2(Z\alpha)}$ & -11.330 \\
higher order QED & 1.23 \\
$\Delta E_Z$ & -56.9(9) \\
$\Delta E_R^p$ & 8.43(8) \\
$\Delta E_{\rm pol}$ & 2.0(8) \\
\hline
\vrule width 0pt height 10pt depth 3pt
$\Delta E({\rm HFS})$ & 1420\,405.7(1.7) \\
\hline
\vrule width 0pt height 10pt depth 3pt
experiment & 1420\,405.751\,7667(9) \\
\hline\hline
\end{tabular}
\caption{Contributions to the hyperfine splitting of the ground state in a hydrogen atom. Uncertainty in the Fermi splitting, $E_F$, is determined by the fine structure constant. The second row is the contribution due to the
anomalous magnetic moment of an electron, $a_e$. Contributions of higher orders, $(Z\alpha)^2$, $\alpha(Z\alpha)$, etc. as well as proton finite size corrections, are taken from Eqs. (\ref{qedcorr})-(\ref{pol}).}
\end{center}
\end{table}

The major part of the contributions mentioned above in
Eq.~(\ref{eq:Zemach_at})-(\ref{pol})  may be considered as contact type
interactions, which depend on the value of the squared density of the
nonrelativistic wave function at the electron-proton coalescence point.
Thus they do not require new extensive calculations, the mean values for
the delta function operators can be taken from \cite{KorPRA06}. The main
task is calculation of the relativistic correction term of order
$E_F(Z\alpha)^2$, which may be performed using the nonrecoil limit of the
two center problem. The obtained effective adiabatic potentials are
subsequently averaged over the radial wave function as it was done for the
$m\alpha^6$ order relativistic correction to ro-vibrational energies in
\cite{KorJPB07,KorPRA08}.

The paper is organized as follows. In Sec.~II.A and II.B we use the NRQED
to derive all the spin-dependent interactions of order $\alpha^6 (m/M)$
and the corresponding potentials in the coordinate space. The radiative
corrections of orders $\alpha(Z\alpha)E_F$ and
$\alpha(Z\alpha)^2\ln^2(Z\alpha)E_F$, as well as proton structure effects,
which may be expressed as contact type interactions in the NRQED, are
given in paragraph C. The perturbation formalism used to obtain the energy
corrections is described in paragraph D. In the next two sections, we
describe the calculation of the relativistic corrections of order
$(Z\alpha)^2E_F$. First, for the HFS of the hydrogen ground state
(Sec.~III), where we rederive the well-known Breit correction
\cite{Breit}, providing a useful check of our approach. Then in Sec.~IV
the $\mbox{H}_2^+$ ion case is considered. Finally, numerical results are
given and discussed in Sec.~V.

\vspace{10mm}

\section{NRQED interactions}

In this section we use the NRQED \cite{Cas86} to describe the
interactions, which are of relevance to our problem. A nice and
illuminative introduction to the NRQED approach may be found in
\cite{Kin96}. The units $c=\hbar=1$ and $e^2=\alpha$ are used in this
section, the elementary charge, $e$, is positive. We consider the low
energy scattering, assuming that the momentum of a particle is of order
$Z\alpha$, and we expand the scattering amplitude in terms of $\alpha$ and
$\mathbf{p}^2$.

\subsection{Tree-level interactions of order $m\alpha^6(m/M)$.}

The momentum 4-vectors for the scattering of an electron (or proton) by
the field of a static external source obey:
\[
p_0'=p_0^{}=E, \qquad \mathbf{q}=\mathbf{p}'-\mathbf{p},\qquad q_0=0,
\qquad q^2=-\mathbf{q}^2,
\]
where $p$ and $p'$ are 4-moments of incident and scattered particles,
respectively.

On-shell Dirac spinors can be presented via the Schr\"odinger-Pauli
spinors as follows
\[
\begin{array}{@{}l}
\displaystyle
u(p)=\sqrt{\frac{E_p+m}{2E_p}}
\left(\begin{matrix}
          \>X\\[1mm]
          \frac{\boldsymbol{\sigma}^P\mathbf{p}}{E_p+m}\,X
      \end{matrix}\right), \qquad
u^*(p)u(p)=(X^*X) = 1,\\[3mm]
\displaystyle\hspace{35mm}
E_p=\sqrt{m^2+\mathbf{p}^2}= m\left(1+\frac{\mathbf{p}^2}{2m^2}
                 -\frac{\mathbf{p}^4}{8m^4}
                 +\frac{\mathbf{p}^6}{16m^6}+\dots\right).
\end{array}
\]
here $\sigma^P_i$ are the two component Pauli matrices and $X$ are the
two-component Schr\"odinger-Pauli wave functions. We assume that Dirac
spinors are normalized as $(u^*u)=1$. That corresponds to the
nonrelativistic normalization: the probability to discover a particle in a
unit of volume is equal to unity. With this normalization Dirac spinors are expanded in
the low-energy limit as follows
\[
u(p)\approx
\left(\begin{matrix} \left[1-\frac{\mathbf{p}^2}{8m^2}
                            +\frac{11\mathbf{p}^4}{128m^4}\right]X\\[2mm]
          \frac{\boldsymbol{\sigma}^P\mathbf{p}}{2m}
          \left[1-\frac{3\mathbf{p}^2}{8m^2}\right]X
      \end{matrix}\right).
\]

The {\em nonrelativistic} scattering amplitude at tree-level for a
scalar static field is determined by the following expansion
\begin{equation}\label{eq:scalarF}
\begin{array}{@{}r@{\;}l}
\displaystyle
A_E(p,p') &\displaystyle
=X^*(\mathbf{p}')(eZ)A_0(\mathbf{q})\biggl(1-\frac{\mathbf{q}^2}{8m^2}
       +i\frac{\boldsymbol{\sigma}^P[\mathbf{q}\times\mathbf{p}]}{4m^2}
       +\frac{3\mathbf{q}^2(p'^2+p^2)}{64m^4}\\[3mm]&
\displaystyle\hspace{27mm}
       +\frac{5(p'^2-p^2)^2}{128m^4}
       -i\frac{3\boldsymbol{\sigma}^P[\mathbf{q}\times\mathbf{p}]
                                                   (p'^2+p^2)}{32m^4}
       +\dots\biggr)X(\mathbf{p})
\end{array}
\end{equation}
and for a vector static field one obtains
\begin{equation}\label{eq:vectorF}
\begin{array}{@{}r@{\;}l}
\displaystyle
A_M(p,p') &\displaystyle
=X^*(\mathbf{p}') (eZ)\mathbf{A}(\mathbf{q})
       \biggl(
          -\frac{\mathbf{p}'+\mathbf{p}}{2m}
          -i\frac{[\boldsymbol{\sigma}^P\times\mathbf{q}]}{2m}
          +\frac{\mathbf{p}'(3p'^2+p^2)+\mathbf{p}(3p^2+p'^2)}{16m^3}
\\[3mm]&\displaystyle\hspace{27mm}
          +i\frac{[\boldsymbol{\sigma}^P\times\mathbf{p}'](3p'^2+p^2)
              -[\boldsymbol{\sigma}^P\times\mathbf{p}](3p^2+p'^2)}{16m^3}
          +\dots
       \biggr)X(\mathbf{p})
\end{array}
\end{equation}
where $Z$ is the charge of a particle $e$, $p$,\dots For an electron
$Z=-1$.

In what follows an index $a$ or $b=1,2$ denotes nucleus 1 or 2 in the
$\mbox{H}_2^+$ ion, indices $i,j=1,3$ are Cartesian coordinates. The
imaginary unit is denoted by upright $\mathrm{i}$.

The higher order vertices of tree level diagrams produce new interactions
($\mathbf{q}=\mathbf{p}'_e\!-\!\mathbf{p}_e$):

a) via Coulomb photon exchange:
\begin{equation}\label{eq:C_exch}
\mathcal{V}_1 =
     e^2\left(
        \mathrm{i}\frac{3\boldsymbol{\sigma}_e^P[\mathbf{q}\times\mathbf{p}_e]
                                              (p'^2_e+p_e^2)}{32m_e^4}
     \right)
     \frac{1}{\mathbf{q}^2}\;(Z_a)
\end{equation}

b) via transverse photon exchange:
\begin{subequations}
\begin{equation}
\mathcal{V}_2 =
     e^2\left(
        -\mathrm{i}\frac{[\boldsymbol{\sigma}^P_e\!\times\!\mathbf{p}'_e]
                                               (3p'^2_e\!+\!p^2_e)
              -[\boldsymbol{\sigma}^P_e\!\times\!\mathbf{p}_e]
                                               (3p^2_e\!+\!p'^2_e)}
                                                        {16m^3_e}
     \right)^i
     \left[-\frac{1}{\mathbf{q}^2}
        \left(\delta^{ij}-\frac{q^iq^j}{\mathbf{q}^2}\right)\right]\;
     \left(-Z_a\frac{\mathbf{P}_a'+\mathbf{P}_a}{2M_a}\right)^j
\end{equation}
\vspace{-4mm}
\begin{equation}
\mathcal{V}_3 =
     e^2\left(
        -\frac{\mathbf{p}'_e(3p'^2_e+p^2_e)+\mathbf{p}_e(3p^2_e+p'^2_e)}{16m^3_e}
     \right)^i
     \;\left[-\frac{1}{\mathbf{q}^2}
        \left(\delta^{ij}-\frac{q^iq^j}{\mathbf{q}^2}\right)\right]\;
     \left(
        -\mathrm{i}Z_a\frac{\boldsymbol{\sigma}_a^P\times(-\mathbf{q})}{2M_a}
     \right)^j
\end{equation}
\vspace{-4mm}
\begin{equation}
\mathcal{V}_4 =
     e^2\left(
        -\mathrm{i}\frac{[\boldsymbol{\sigma}^P_e\!\times\!\mathbf{p}'_e]
                                                  (3p'^2_e\!+\!p^2_e)
              -[\boldsymbol{\sigma}^P_e\!\times\!\mathbf{p}_e]
                                                  (3p^2_e\!+\!p'^2_e)}
                                                     {16m^3_e}
     \right)^i
     \left[-\frac{1}{\mathbf{q}^2}
        \left(\delta^{ij}-\frac{q^iq^j}{\mathbf{q}^2}\right)\right]\;
     \left(
        -\mathrm{i}Z_a\frac{\boldsymbol{\sigma}_a^P\times(-\mathbf{q})}{2M_a}
     \right)^j
\end{equation}
\end{subequations}
In parentheses here are the vertex functions of the effective NRQED
interaction taken from Eqs.~(\ref{eq:scalarF})-(\ref{eq:vectorF}). The
approximate transverse photon propagator (see \cite{BLP}, \S\ 83) is placed
in the square brackets.

The obtained potentials can be simplified as follows:
\[
\mathcal{V}_2 = e^2
     \left(
        -\mathrm{i}\frac{2[\boldsymbol{\sigma}_e^P\!\times\!\mathbf{q}]
                                               (p'^2_e\!+\!p^2_e)
               +[\boldsymbol{\sigma}_e^P\!\times\!(\mathbf{p}'_e+\mathbf{p}_e)]
                                               (p'^2_e\!-\!p^2_e)}{16m_e^3}
     \right)^i
     \left[\frac{1}{\mathbf{q}^2}
        \left(\delta^{ij}-\frac{q^iq^j}{\mathbf{q}^2}\right)\right]\;
     \left(Z_a\frac{\mathbf{P}_a}{M_a}\right)^j
\]
The last term in the first brackets produces a symmetric operator with the
property: $(\varphi,A\varphi)=0$, for an arbitrary $\varphi$. That means
that this operator is identical to zero operator, and $\mathcal{V}_2$
may be rewritten
\begin{subequations}
\begin{equation}
\mathcal{V}_2 = -e^2
     \left(
        \mathrm{i}\frac{[\boldsymbol{\sigma}_e^P\!\times\!\mathbf{q}]
                                               (p'^2_e\!+\!p^2_e)}{8m_e^3}
     \right)^i
     \frac{1}{\mathbf{q}^2}
     \left(Z_a\frac{\mathbf{P}_a}{M_a}\right)^i
\end{equation}
In a similar way the other operators may be simplified:
\begin{equation}
\mathcal{V}_3 =
     e^2\left(
        \frac{\mathbf{p}_e(p_e'^2+p_e^2)}{4m_e^3}
     \right)^i
     \frac{1}{\mathbf{q}^2}
     \left(
        \mathrm{i}\,Z_a\frac{\boldsymbol{\sigma}_a^P\!\times\!\mathbf{q}}{2M_a}
     \right)^i
\end{equation}
\vspace{-4mm}
\begin{equation}
\mathcal{V}_4 =
     -e^2\left(
        \frac{[\boldsymbol{\sigma}_e^P\!\times\!\mathbf{q}]
                                               (p_e'^2\!+\!p_e^2)}{8m_e^3}
     \right)^i
     \frac{1}{\mathbf{q}^2}
     \left(
        Z_a\frac{\boldsymbol{\sigma}_a^P\!\times\!\mathbf{q}}{2M_a}
     \right)^i
\end{equation}
\end{subequations}

Transforming potentials $\mathcal{V}_n$ to the coordinate space
($\mathbf{r}_a=\mathbf{R}_e-\mathbf{R}_a$, where $\mathbf{R}_e$,
$\mathbf{R}_a$ are the coordinates of electron and nuclei with respect to
the center of mass) one gets, using the notation $\left\{A,B\right\}=AB+BA$:
\begin{equation}\label{eq:tree_space}
\begin{array}{@{}l}
\displaystyle
\mathcal{V}_1 =
   -\alpha\frac{3Z_a}{16m_e^4}
      \left\{
         p_e^2,\frac{1}{r_{a}^3}[\mathbf{r}_{a}\!\times\!\mathbf{p}_e]
      \right\}\mathbf{s}_e,
\\[4mm]\displaystyle
\mathcal{V}_2 =
   \alpha\frac{Z_a}{4m_e^3M_a}
      \left\{
         p_e^2,\frac{1}{r_{a}^3}[\mathbf{r}_{a}\!\times\!\mathbf{P}_a]
      \right\}\mathbf{s}_e,
\\[4mm]\displaystyle
\mathcal{V}_3 =
   -\alpha\frac{1}{2m_e^3}
      \left\{
         p_e^2,\frac{1}{r_{a}^3}[\mathbf{r}_{a}\!\times\!\mathbf{p}_e]
      \right\}\boldsymbol{\mu}_a,
\\[4mm]\displaystyle
\mathcal{V}_4 =
   -\alpha\frac{1}{4m_e^3}
      \left\{
         p_e^2,
         \left[
            \frac{8\pi}{3}\mathbf{s}_e\boldsymbol{\mu}_a
                                                \delta(\mathbf{r}_{a})
            -\frac{r_{a}^2\mathbf{s}_e\boldsymbol{\mu}_a
                 \!-\!3(\mathbf{s}_e\mathbf{r}_{a})
                       (\boldsymbol{\mu}_a\mathbf{r}_{a})}{r_{a}^5}
      \right]
      \right\}.
\end{array}
\end{equation}
Here $\boldsymbol{\mu}_a$ is the magnetic moment operator for nucleus $a$.
Only $\mathcal{V}_4$ involves both electron and nuclear spins and contributes to $b_F$.

\begin{figure}[t]
\includegraphics*[width=0.75\textwidth]{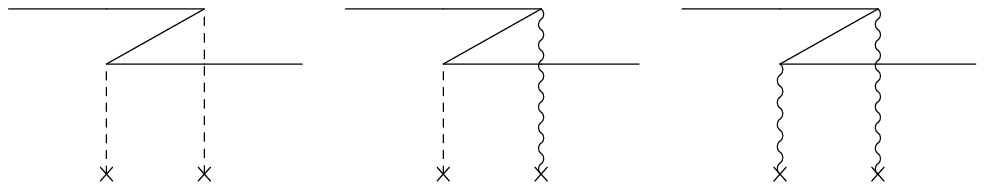}\\[3mm]
\includegraphics*[width=0.20\textwidth]{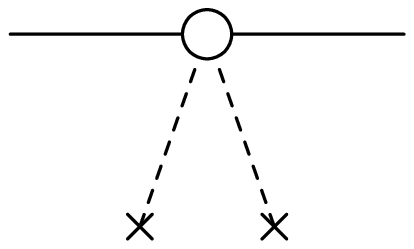}
\hspace{9mm}
\raisebox{-3.4pt}{\includegraphics[width=0.20\textwidth]{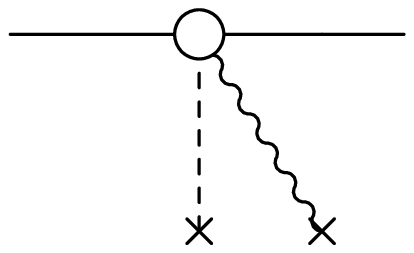}}
\hspace{9mm}
\includegraphics*[width=0.20\textwidth]{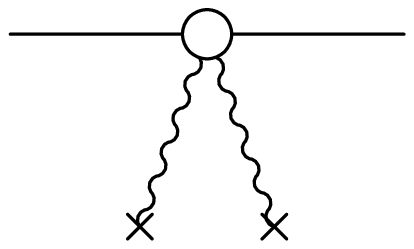}\\[3mm]
\hspace{4mm}
\begin{tabular}{c@{\hspace{33mm}}c@{\hspace{33mm}}c}
$\displaystyle\frac{e^2\mathbf{E}^2}{8m^3}$ &
$\displaystyle
  \frac{e^2\boldsymbol{\sigma}\cdot[\mathbf{A}\times\mathbf{E}]}{4m^2}$ &
$\displaystyle\frac{e^2\mathbf{A}^2}{2m}$ \\[2mm]
$\alpha^6$ & $\alpha^6\left(\frac{m}{M}\right)$ &
$\alpha^6\left(\frac{m}{M}\right)^2$
\end{tabular}
\caption{Top row: QED scattering diagrams. Bottom row: NRQED
seagull interactions inferred from QED diagrams.}\label{fig:seagull}
\end{figure}

\subsection{Seagull-type interactions}

In Fig.~\ref{fig:seagull}, three NRQED seagull diagrams are presented.
They may be obtained from the corresponding QED $Z$-diagrams by expanding
the scattering amplitude in terms of $\mathbf{p}^2$. The double Coulomb
photon exchange diagram has a leading order $\alpha^6$, however it does
not involve interactions dependent on spin. The third diagram is double
transverse photon exchange has a recoil order $(m/M)^2$, and is also out
of interest for present consideration.

The potentials which stems from the seagull vertex with one Coulomb and one
transverse photon lines can be expressed as follows
($\mathbf{q}_1=\mathbf{P}_1-\mathbf{P}'_1$,
$\mathbf{q}_2=\mathbf{P}_2-\mathbf{P}'_2$):
\begin{subequations}
\begin{equation}
\mathcal{V}_5 =
     e^4\frac{\boldsymbol{\sigma}_e^P}{4m_e^2}
     \left[
        \left\{
           \left[
              -\frac{1}{\mathbf{q}_b^2}
              \left(\delta^{ij}-\frac{q_b^iq_b^j}{\mathbf{q}_b^2}\right)
           \right]
           \left(-Z_b\frac{\mathbf{P}'_b+\mathbf{P}_b}{2M_b}\right)^i
        \right\}
        \times
        \mathrm{i}\left[\frac{\mathbf{q}_a}{\mathbf{q}_a^2}\right]^j(Z_a)
     \right]
\end{equation}
\vspace{-4mm}
\begin{equation}
\mathcal{V}_6 =
     e^4\frac{\boldsymbol{\sigma}_e^P}{4m_e^2}
     \left[
        \left\{
           \left[
              -\frac{1}{\mathbf{q}_b^2}
              \left(\delta^{ij}-\frac{q_b^iq_b^j}{\mathbf{q}_b^2}\right)
           \right]
           \left(
              -\mathrm{i}
                 Z_b\frac{\boldsymbol{\sigma}_b^P\times(-\mathbf{q}_b)}{2M_b}
           \right)^i
        \right\}
        \times
        \mathrm{i}\left[\frac{\mathbf{q}_a}{\mathbf{q}_a^2}\right]^j(Z_a)
     \right]
\end{equation}
\end{subequations}
The sources $Z_a$ and $Z_b$ may belong to a same particle or to two
different particles.

In the coordinate space one has, for $a \neq b$,
\begin{subequations}\label{eq:seagull_space}
\begin{equation}
\mathcal{V}_5=-\alpha^2\frac{Z_aZ_b}{4m_e^2M_b}
    \left\{
       \frac{[\mathbf{r}_a\!\times\!\mathbf{P}_b]}{r_a^3r_b}
       +\frac{[\mathbf{r}_a\!\times\!\mathbf{r}_b]
              (\mathbf{r}_b\mathbf{P}_b)}{r_a^3r_b^3}
    \right\}\mathbf{s}_e\,,
\end{equation}
\vspace{-4mm}
\begin{equation}
\mathcal{V}_6=
    \alpha^2\frac{Z_aZ_b\mu_b}{2m_e^2M_b}
    \frac{[\mathbf{r}_a\!\times\!\mathbf{s}_e]
          [\mathbf{r}_b\!\times\!\mathbf{s}_b]}{r_a^3r_b^3} =
    \alpha^2\frac{Z_a}{2m_e^2}
    \frac{[\mathbf{r}_a\!\times\!\mathbf{s}_e]
          [\mathbf{r}_b\!\times\!\boldsymbol{\mu}_b]}{r_a^3r_b^3}\,.
\end{equation}
Using $[\mathbf{a}\!\times\!\mathbf{b}][\mathbf{c}\!\times\!\mathbf{d}]=
(\mathbf{ac})(\mathbf{bd})\!-\!(\mathbf{cb})(\mathbf{ad})$, one may
further simplify $\mathcal{V}_6$.

When the sources
coincide ($a=b$), the interactions modify as follows:
\begin{equation}
\mathcal{V}_7=-\alpha^2\frac{Z_a^2}{4m_e^2M_a}
     \frac{[\mathbf{r}_a\!\times\!\mathbf{P}_a]}{r_a^4}\mathbf{s}_e\,,
\end{equation}
\vspace{-4mm}
\begin{equation}
\mathcal{V}_8=\alpha^2\frac{Z_a}{2m_e^2}
    \frac{[\mathbf{r}_a\!\times\!\mathbf{s}_e]
          [\mathbf{r}_a\!\times\!\boldsymbol{\mu}_a]}{r_a^6}
  = -\alpha^2\frac{Z_a}{2m_e^2}
    \left[
       \frac{(\mathbf{r}_a\mathbf{s}_e)
             (\mathbf{r}_a\boldsymbol{\mu}_a)
             -(1/3)r_a^2(\mathbf{s}_e\boldsymbol{\mu}_a)}{r_a^6}
       -\frac{2}{3}\frac{(\mathbf{s}_e\boldsymbol{\mu}_a)}{r_a^4}
    \right].
\end{equation}
\end{subequations}
Among those terms, only $\mathcal{V}_6$ and $\mathcal{V}_8$ contribute to
$b_F$.

\subsection{Contact type NRQED interactions.}

Here we introduce corrections already mentioned in the introduction, which
enter into the NRQED Lagrangian as contact type interactions, since they
reproduce effects of the relativistic scale.

\begin{itemize}
\item Radiative interactions of order $\alpha(Z\alpha)E_F$:
\begin{equation}\label{eq:aZa}
V_r^{(6)}
 = \alpha^3\frac{8\pi Z}{3m_e}\frac{\mu_p}{M_p}
      \left(\ln{2}-\frac{13}{4}+\frac{3}{4}\right)
      \bigl(\mathbf{s}_e\cdot\mathbf{I}_a\bigr)\delta(\mathbf{r}_a),
\end{equation}
and of order $\alpha(Z\alpha)^2\ln^2(Z\alpha)E_F$
\begin{equation}
V_r^{(7)}
 = \alpha^4\ln^2(Z\alpha)\frac{8\pi Z^2}{3m_e}\frac{\mu_p}{M_p}
      \left(-\frac{8}{3\pi}\right)
      \bigl(\mathbf{s}_e\cdot\mathbf{I}_a\bigr)\delta(\mathbf{r}_a).
\end{equation}

\item Zemach term ($(Z\alpha)(m/\Lambda)E_F$):
\begin{equation}\label{eq:Zemach}
V_Z
 = -2\alpha R_Z(1+\delta^{\rm rad}_Z)\,
      \frac{8\pi Z}{3m_e}\frac{\mu_p}{M_p}
      \>\bigl(\mathbf{s}_e\cdot\mathbf{I}_a\bigr)\delta(\mathbf{r}_a).
\end{equation}
where $R_Z=1.045(16)$ \cite{Shabaev}.

\item Recoil correction of order $(Z\alpha)(m/M)E_F$
\cite{BodYen88,Shabaev}:
\begin{equation}
V_{recoil}
 = \alpha\left[5.48(6)\cdot10^{-6}\right]\frac{8\pi}{3m_e}\frac{\mu_p}{M_p}
      \>\bigl(\mathbf{s}_e\cdot\mathbf{I}_a\bigr)\delta(\mathbf{r}_a).
\end{equation}
The correction of order $(Z\alpha)^2(m/M)E_F$ from Eq.~(\ref{eq:recoil}) has
been omitted since it has to be considered in the context of higher order
corrections, which are out of the scope of present consideration.
\item proton polarizability \cite{Fau02}:
\begin{equation}\label{eq:pol}
V_{pol}
 = \alpha\left[1.4(6)\cdot10^{-6}\right]\frac{8\pi}{3m_e}\frac{\mu_p}{M_p}
      \>\bigl(\mathbf{s}_e\cdot\mathbf{I}_a\bigr)\delta(\mathbf{r}_a).
\end{equation}
\end{itemize}

\subsection{Perturbation formalism.}

To calculate the bound state problem we use the nonrelativistic
Rayleigh-Schr\"odinger perturbation theory, where the starting point, the
zero order approximation, is the nonrelativistic Schr\"odinger equation:
\begin{equation}
H_0\Psi_0=E_0\Psi_0,
\end{equation}
and the perturbation is the effective Hamiltonian $H_{\rm eff}$ derived
from the NRQED Lagrangian and
\begin{equation}\label{eq:perturb}
\Delta E = \left\langle H_{\rm eff} \right\rangle
   + \left\langle H_{\rm eff}Q(E_0-H_0)^{-1}QH_{\rm eff} \right\rangle
   + \left\langle H_{\rm eff} \right\rangle
     \left\langle \left[\frac{\partial}{\partial E} H_{\rm eff}\right]\right\rangle
   + \dots
\end{equation}
where $Q$ is a projector operator on the subspace orthogonal to the
zero-order wave function. $H_{\rm eff}$ has contributions of different
orders in $\alpha$:
\[
H_{\rm eff} = H^{(4)}+H^{(5)}+H^{(6)}+\dots
\]
Then for our case the complete contribution at order of $\alpha^6(m/M)$ to
the hyperfine structure of hydrogen atom and ion can be expressed by
\begin{equation}\label{eq:pert}
\Delta E_{\rm hfs} = \left\langle H^{(6)} \right\rangle+
\left\langle H_u^{(4)}Q(E_0-H_0)^{-1}QH_v^{(4)}\right\rangle.
\end{equation}
where $H_u^{(4)}$ and $H_v^{(4)}$ are parts of the Breit-Pauli Hamiltonian
taken so that the second term in (\ref{eq:pert}) contributes to that
particular order. Since the effective Hamiltonian $H^{(6)}$ for the HFS
does not depend on $E$ explicitly, the last term of Eq.~(\ref{eq:perturb})
vanishes. In the following, the second-order contribution and the
first-order contribution $\langle H^{(6)} \rangle$ will be denoted $\Delta
E_A$ and $\Delta E_B$, respectively.

\section{HFS in the hydrogen ground state}

In the remaining part of this work we will be using the atomic units
($e=\hbar=1$ and $c=\alpha^{-1}$).
First, we consider the case of the HFS of the ground state of a hydrogen
atom. Our derivation is somewhat similar to the one done by Nio and Kinoshita
in \cite{Kin97}. The divergent part is, however, treated in a different
way by explicitly separating out and cancelling the divergences.
We start from the nonrelativistic Schr\"odinger equation:
\begin{equation}\label{Hground}
(H_0-E_0)\Psi_0 = \left(\frac{p_e^2}{2m_e}+V\right)\Psi_0,
\end{equation}
where
\begin{equation}
V=-\frac{Z}{r}.
\end{equation}

\subsection{Separating divergences in the $m\alpha^6(m/M)$ order effective Hamiltonian}

The effective Hamiltonian of order $m\alpha^6(m/M)$ is obtained from
Eqs.~(\ref{eq:tree_space}) and (\ref{eq:seagull_space}),  interactions
$\mathcal{V}_4$ and $\mathcal{V}_8$, expressed in atomic units.
It has the form:
\begin{equation}\label{eq:eff_Ham}
\begin{array}{@{}r@{\;}l}
H^{(6)} & \displaystyle
 = \alpha^4\frac{2}{3m_e}
   \left[
      -\frac{1}{4m_e^2}\Bigl\{p_e^2,4\pi\delta(\mathbf{r})\Bigr\}
      +\frac{Z}{2m_e}\;\frac{1}{r^4}
   \right]
   \mathbf{s}_e\boldsymbol{\mu}_p
\end{array}
\end{equation}
where $\boldsymbol{\mu}_p=(\mu_p/M_p)\,\mathbf{I}_p$,
then using the relation
\begin{equation}
\left\langle Z^2/r^4 \right\rangle =
  -\left\langle
     \left(\mathbf{p}_eV\right)^2
  \right\rangle
  =
  -\left\langle
     \frac{p_e^2V^2+V^2p_e^2}{2}
  \right\rangle
  +\left\langle \mathbf{p}_e V^2\mathbf{p}_e \right\rangle
  -4\pi Z\left\langle V\delta(\mathbf{r})\right\rangle
\end{equation}
obtained by integration by parts,
and equation $p_e^2\Psi_0=2(E_0-V)\Psi_0$, one gets:
\begin{equation}\label{eq:eff_Ham_b}
\Delta E^{(6)}_B = \left\langle H^{(6)}\right\rangle
 = \alpha^4\frac{2}{3Zm_e}\frac{\mu_p}{M_p}
    \biggl[
      \frac{2\pi Z\left\langle V\delta(\mathbf{r}) \right\rangle}{m_e}
      +\bigl\langle V^3 \bigr\rangle
      -E_0\bigl\langle V^2 \bigr\rangle
      +\frac{\bigl\langle \mathbf{p}_eV^2\mathbf{p}_e \bigr\rangle}{2m_e}
      -\frac{4\pi ZE_0\left\langle \delta(\mathbf{r}) \right\rangle}{m_e}
    \biggr]
    \left\langle\mathbf{s}_e\!\cdot\!\mathbf{I}_p\right\rangle.
\end{equation}
In Eq.~(\ref{eq:eff_Ham_b}), the divergent contributions are now
explicitly collected in the first two terms.

\subsection{Separating divergences in the second order contribution}

The second order contribution of order $m\alpha^6(m/M)$ to the spin-spin
interaction can be easily identified from various combinations of
terms of the Breit-Pauli Hamiltonian and may be written:
\begin{equation}\label{eq:2order}
\Delta E_A^{(6)} =
   2\alpha^4
   \left\langle
      -\frac{\mathbf{p}_e^4}{8m_e^3}
      +\frac{Z}{8m_e^2}\,4\pi\delta(\mathbf{r})
      \bigg|Q(E_0-H_0)^{-1}Q\bigg|
      \frac{8\pi}{3m_e}\;
         \mathbf{s}_e\boldsymbol{\mu}_p\,\delta(\mathbf{r})
   \right\rangle,
\end{equation}
This contribution is divergent due to presence of the delta-function
operators on both sides of the second order iteration.

Let us consider the two operators:
\[
H_B^{(1)}=4\pi\delta(\mathbf{r}),
\qquad
H_B^{(2)}=-\frac{\mathbf{p}_e^4}{8m_e^3}
          +\frac{Z}{8m_e^2}\,4\pi\delta(\mathbf{r}),
\]
and introduce the wavefunction $\Psi_B^{(1)}$ solution
 of equation
\begin{equation}\label{WFdelta}
(E_0-H_0)\Psi_B^{(1)}=
Q\left[4\pi\,\delta(\mathbf{r})\right]\Psi_0=QH_B^{(1)}\Psi_0.
\end{equation}
$\Psi_B^{(1)}$ behaves as $1/r$ at $r\to0$.
We introduce a less singular function $\tilde{\Psi}_B^{(1)}$
defined by
\begin{equation}\label{eq:1st_order_wf1}
\Psi_B^{(1)} = -\frac{2m_e\Psi_0}{r}+\tilde{\Psi}_B^{(1)}
             = U_1\Psi_0+\tilde{\Psi}_B^{(1)},
\end{equation}
where
$U_1 = -\frac{2m_e}{r} = \frac{2m_e}{Z}V$.
The function $\tilde{\Psi}_B^{(1)}$ behaves as $\ln{r}$ at $r\to0$.
It satisfies equation
\begin{equation}\label{eq:HB1}
(E_0-H_0)\tilde{\Psi}_B^{(1)} =
   \left({H'_{B\!}}^{(1)}
   -\left\langle{H'_{B\!}}^{(1)}\right\rangle\right)\Psi_0,
\end{equation}
where
\begin{equation}\label{eq:def:HB1'}
{H'_{B\!}}^{(1)}=-(E_0-H_0)U_1-U_1(E_0-H_0)+H_B^{(1)}.
\end{equation}
Similar computations can be applied to the scalar part of the
Breit-Pauli Hamiltonian $H_B^{(2)}$:
\begin{equation}\label{WFbreit}
\begin{array}{@{}l}
\displaystyle
(E_0-H_0)\Psi_B^{(2)}=
Q\left[-\frac{p^4}{8m_e^3}
       +\frac{Z\pi}{2m_e^2}\delta(\mathbf{r})\right]\Psi_0
\\[4mm]\displaystyle
\Psi_B^{(2)} = \frac{Z\Psi_0(r)}{4m_er}+\tilde{\Psi}_B^{(2)}
             = U_2\Psi_0+\tilde{\Psi}_B^{(2)},
\qquad
U_2 = \frac{Z}{4m_er} = -\frac{1}{4m_e}V.
\end{array}
\end{equation}
and
\begin{equation}\label{eq:HB2}
{H'^{(2)}_{B\!}}=-(E_0-H_0)U_2-U_2(E_0-H_0)+H_B^{(2)}.
\end{equation}
Using systematically that
\[
\left\langle
   \Psi_0\Big|H_B^{(2)}Q(E_0-H_0)^{-1}QH_B^{(1)}\Big|\Psi_0
\right\rangle =
\left\langle
   \Psi_0\Big|H_B^{(2)}Q\Big|\Psi_B^{(1)}
\right\rangle,
\qquad
\mbox{or}\quad
\left\langle
   \Psi_B^{(2)}\Big|QH_B^{(1)}\Big|\Psi_0
\right\rangle,
\]
one may separate the divergent singularities in the following way:
\begin{equation}\label{eq:transf}
\begin{array}{@{}r@{\;}l}
\Delta E_A & = \displaystyle
   \alpha^4\frac{4\mu_p}{3m_eM_p}\;
   \left\langle\mathbf{s}_e\!\cdot\!\mathbf{I}_p\right\rangle
   \left\langle
      \Psi_0\Big|H_B^{(2)}Q(E_0-H_0)^{-1}QH_B^{(1)}\Big|\Psi_0
   \right\rangle
\\[3mm] & \displaystyle
 = \alpha^4\frac{2\mu_p}{3m_eM_p}\;
   \left\langle\mathbf{s}_e\!\cdot\!\mathbf{I}_p\right\rangle
\left(
   2\left\langle
      \Psi_0\Big|\left(H_B^{(2)}
         -\left\langle H_B^{(2)}\right\rangle\right)U_1\Big|\Psi_0
   \right\rangle+
   2\left\langle
      \Psi_0\Big|\left(H_B^{(2)}
         -\left\langle H_B^{(2)}\right\rangle\right)\Big|\tilde{\Psi}_B^{(1)}
   \right\rangle
\right)
\\[3mm] & \displaystyle
 = \alpha^4\frac{2\mu_p}{3m_eM_p}\;
   \left\langle\mathbf{s}_e\!\cdot\!\mathbf{I}_p\right\rangle
\Bigl(
   2\left\langle
      \Psi_0\Big|\left(H_B^{(2)}
         -\left\langle H_B^{(2)}\right\rangle\right)U_1\Big|\Psi_0
   \right\rangle+
   2\left\langle
      \Psi_0\Big|H_B^{(2)}Q(E_0-H_0)^{-1}Q{H'_{B\!}}^{(1)}\Big|\Psi_0
   \right\rangle
\Bigr)
\\[3mm] & \displaystyle
 = \alpha^4\frac{2\mu_p}{3m_eM_p}\;\
   \left\langle\mathbf{s}_e\!\cdot\!\mathbf{I}_p\right\rangle
\Bigl(
   2\left\langle
      \Psi_0\Big|\left(H_B^{(2)}
         -\left\langle H_B^{(2)}\right\rangle\right)U_1\Big|\Psi_0
   \right\rangle+
   2\left\langle
      \Psi_0\Big|U_2({H'_{B\!}}^{(1)}
      -\langle{H'_{B\!}}^{(1)}\rangle)\Big|\Psi_0
   \right\rangle
\\[3mm] & \displaystyle \hspace{45mm}
   +2\left\langle
      \Psi_0\Big|{H'_{B\!}}^{(2)}Q(E_0-H_0)^{-1}Q{H'_{B\!}}^{(1)}\Big|\Psi_0
   \right\rangle
\Bigr)
\end{array}
\end{equation}
The first two terms of the last expression may be rewritten as the average of
a new effective Hamiltonian contributing to the $m\alpha^6(m/M)$ order:
\begin{equation}
\begin{array}{@{}r@{\;}l}
\displaystyle
H'^{(6)} & \displaystyle
 = \alpha^4\frac{2\mu_p}{3m_eM_p}
   \biggl\{
      \left( H_B^{(2)}U_1+U_1H_B^{(2)}\right)
      +\left( H_B^{(1)}U_2+U_2H_B^{(1)}\right)
      -2\left\langle H_B^{(2)}\right\rangle U_1
      -2\left\langle H_B^{(1)}\right\rangle U_2
\\[2mm] & \hspace{65mm}
      -U_1(E_0-H_0)U_2-U_2(E_0-H_0)U_1
   \biggr\}
   \left\langle\mathbf{s}_e\!\cdot\!\mathbf{I}_p\right\rangle.
\end{array}
\end{equation}
Using regularization and integration by parts in a similar way as in
Appendix B of \cite{KorJPB07} its expectation value may be finally written
in the form
\begin{equation}\label{eq:new_eff_Ham}
\begin{array}{@{}r@{\;}l}
\displaystyle
\left\langle H'^{(6)}\right\rangle & \displaystyle
 = \alpha^4\frac{2\mu_p}{3Zm_eM_p}
   \biggl[
      -\frac{2\pi Z\left\langle V\delta(\mathbf{r}) \right\rangle}{m_e}
      -\left\langle V^3\right\rangle
      +\frac{\left\langle \mathbf{p}V^2\mathbf{p} \right\rangle}{2m_e}
      +3E_0\left\langle V^2\right\rangle
      -2E_0^2\left\langle V\right\rangle
\\[1mm] & \hspace{75mm} \displaystyle
      -4m_e\left\langle H_B^{(2)}\right\rangle \langle V \rangle
      +\frac{Z\bigl\langle H_B^{(1)}\bigr\rangle \langle V \rangle}{2m_e}
   \biggr]
   \left\langle\mathbf{s}_e\!\cdot\!\mathbf{I}_p\right\rangle.
\end{array}
\end{equation}
All the divergent terms of Eq.~(\ref{eq:transf}) are collected as the first two
terms of Eq.~(\ref{eq:new_eff_Ham}). They clearly cancel out those of
Eq.~\ref{eq:eff_Ham_b}.

The remaining part of the second order iteration contribution (the last
term in Eq.~(\ref{eq:transf})) is finite
\begin{equation}\label{eq:H4p_iter}
\Delta E'^{(6)}_A =
   \alpha^4\frac{4\mu_p}{3m_eM_p}
   \left\langle
      \Psi_0\Big|{H'_{B\!}}^{(2)}Q(E_0-H_0)^{-1}Q{H'_{B\!}}^{(1)}\Big|\Psi_0
   \right\rangle
   \left\langle\mathbf{s}_e\!\cdot\!\mathbf{I}_p\right\rangle,
\end{equation}
where ${H'_{B\!}}^{(1)}$ and ${H'_{B\!}}^{(2)}$ are defined above in
Eqs.~(\ref{eq:def:HB1'}) and (\ref{eq:HB2}), respectively.

Summing up $\left\langle H^{(6)}\right\rangle$ and $\left\langle
H'^{(6)}\right\rangle$ from Eqs.~(\ref{eq:eff_Ham_b}) and
(\ref{eq:new_eff_Ham}) one gets a finite expression as well
\begin{equation}\label{eq:H6p}
\begin{array}{@{}l}
\displaystyle
\Delta E'^{(6)}_B =
   \alpha^4\frac{2\mu_p}{3Zm_eM_p}
   \biggl[
      \frac{\left\langle \mathbf{p}V^2\mathbf{p} \right\rangle}{m_e}
      +2E_0\left\langle V^2\right\rangle
      -2E_0^2\left\langle V\right\rangle
      -\frac{ZE_0}{m_e}\>4\pi\left\langle \delta(\mathbf{r}) \right\rangle
\\[3mm]\hspace{70mm}\displaystyle
      -4m_e\left\langle H_B^{(2)}\right\rangle \langle V \rangle
      +\frac{Z\bigl\langle H_B^{(1)}\bigr\rangle \langle V \rangle}{2m_e}
   \biggr]
   \left\langle\mathbf{s}_e\!\cdot\!\mathbf{I}_p\right\rangle.
\end{array}
\end{equation}

\subsection{Calculation of expectation values and final result}

We now check that Eq.~(\ref{eq:H6p}) leads to the usual result
in the case of an 1s hydrogen atom. Here, $\Psi_0=2Z^{3/2}e^{-Zr}$ is the
ground state wave function.
First we look for a solution of equation:
\[
(E_0-H_0)\tilde{\Psi}_B^{(1)} =
   \left(\!
      {H'_{B\!}}^{(1)}\!-\!\left\langle{\!H'_{B\!}}^{(1)}\!\right\rangle
   \right)\Psi_0,
\qquad
{H'_{B\!}}^{(1)}=
   -\frac{2}{r^2}\partial_r\!-2U_1(E_0\!-\!H_0),
\qquad
\left\langle\! H_B^{(1)} \!\right\rangle =4Z^3,
\]
and get
\[
\tilde{\Psi}_B^{(1)} = 4Z(\ln{r}-1+Zr)\Psi_0.
\]
The next step is to calculate the expectation value of
\[
\left\langle \Psi_0\left|
   {H'_{B\!}}^{(2)}Q
\right|\tilde{\Psi}_B^{(1)} \right\rangle =
\left\langle \Psi_0\left|
   {H'_{B\!}}^{(2)}\!-\left\langle{\!H'_{B\!}}^{(2)}\!\right\rangle
\right|\tilde{\Psi}_B^{(1)} \right\rangle,
\]
and to get the finite part of the second order contribution
\[
\Delta E'^{(6)}_A =
   \frac{4Zm_e\alpha^4}{3}\frac{\mu_p}{M_p}
   \left\langle
      \Psi_0\Big|{H'_{B\!}}^{(2)}Q(E_0-H_0)^{-1}Q{H'_{B\!}}^{(1)}\Big|\Psi_0
   \right\rangle
   \left\langle\mathbf{s}_e\!\cdot\!\mathbf{I}_p\right\rangle
   = \frac{3(Z\alpha)^2}{2}E_F
   \left\langle\mathbf{s}_e\!\cdot\!\mathbf{I}_p\right\rangle.
\]

The expectation values of the operators involved in $\Delta E'^{(6)}_B$ are
$\langle\mathbf{p}V^2\mathbf{p}\rangle = Z^2\langle V^2\rangle$,
$\langle V^2\rangle=2Z^2$, and $\langle V\rangle=-Z$. This
contribution is immediately obtained to be 0, indeed
\[
\Delta E'^{(6)}_B =
   \frac{2Zm_e\alpha^4}{3}\frac{\mu_p}{M_p}
   \left[
      2Z^5\!-\!2Z^5\!+\!\frac{Z^5}{2}\!+\!2Z^5\!-\!\frac{Z^5}{2}\!-\!2Z^5
   \right]
   \left\langle\mathbf{s}_e\!\cdot\!\mathbf{I}_p\right\rangle
   = 0.
\]
Thus, the total contribution is
\[
\Delta E^{(6)} = \Delta E'^{(6)}_A+\Delta E'^{(6)}_B =
      \frac{3}{2}(Z\alpha)^2 E_F
   \left\langle\mathbf{s}_e\!\cdot\!\mathbf{I}_p\right\rangle.
\]
that exactly matches the well-known Breit relativistic correction \cite{Breit}.

\section{Hydrogen molecular ion $\mbox{H}_2^+$}

Now we are ready to study the hydrogen molecular ion. As in the previous
section we start from the nonrelativistic equation with the Hamiltonian:
\begin{equation}\label{Hplus_ground}
H_0 = \frac{p^2}{2m_e}+V,\qquad V=-\frac{Z_1}{r_1}-\frac{Z_2}{r_2}.
\end{equation}
We will assume here that $Z_1\!=\!Z_2\!=\!Z$ and
$\boldsymbol{\mu}_1\!=\!(\mu_p/M_p)\mathbf{I}_1$,
$\boldsymbol{\mu}_2\!=\!(\mu_p/M_p)\mathbf{I}_2$, where $\mathbf{I}_1$ and
$\mathbf{I}_2$ are the two proton spin operators.

The second order contribution of the spin-spin interaction of order
$m\alpha^6(m/M)$ is expressed by
\begin{equation}
\Delta E_A =
   2\alpha^4
   \left\langle
      -\frac{\mathbf{p}_e^4}{8m_e^3}
      +\frac{Z}{8m_e^2}\,4\pi
         \left[\delta(\mathbf{r}_1)\!+\!\delta(\mathbf{r}_2)\right]
      \bigg|Q(E_0-H_0)^{-1}Q\bigg|
      \frac{8\pi}{3m_e}\;
         \mathbf{s}_e
         \left[
            \boldsymbol{\mu}_1\delta(\mathbf{r}_1)\!+\!\boldsymbol{\mu}_2\delta(\mathbf{r}_2)
         \right]
   \right\rangle
\end{equation}

The effective Hamiltonian of order $m\alpha^6(m/M)$ is obtained from
Eqs.~(\ref{eq:tree_space}) and (\ref{eq:seagull_space}) of Sec.~II. Now we
have three interactions, $\mathcal{V}_4$, $\mathcal{V}_6$, and
$\mathcal{V}_8$, because we have as well the seagull interaction $\mathcal{V}_6$ with two
different nuclei,
\begin{subequations}
\begin{equation}
\mathcal{V}_{4a} =
   -\alpha^4\frac{1}{4m_e^3}
      \left\{
         p_e^2,
         \left[
            \frac{8\pi}{3}\mathbf{s}_e\boldsymbol{\mu}_a
                                                \delta(\mathbf{r}_a)
            -\frac{r_a^2\mathbf{s}_e\boldsymbol{\mu}_a
                 \!-\!3(\mathbf{s}_e\mathbf{r}_a)
                       (\boldsymbol{\mu}_a\mathbf{r}_a)}{r_a^5}
      \right]
      \right\},
\end{equation}
\vspace{-4mm}
\begin{equation}
\mathcal{V}_6=
    \alpha^4\frac{Z}{6m_e^2}
    \left\{
    \frac{2(\mathbf{r}_1\mathbf{r}_2)(\mathbf{s}_e\boldsymbol{\mu}_I)}
                                                         {r_1^3r_2^3}
    +\frac{(\mathbf{r}_1\mathbf{r}_2)(\mathbf{s}_e\boldsymbol{\mu}_I)
          \!-\!3(\mathbf{r}_1\mathbf{s}_e)(\mathbf{r}_2\boldsymbol{\mu}_2)
          \!-\!3(\mathbf{r}_2\mathbf{s}_e)(\mathbf{r}_1\boldsymbol{\mu}_1)}
                                                       {r_1^3r_2^3}
    \right\},
\end{equation}
\vspace{-4mm}
\begin{equation}
\mathcal{V}_{8a} = \alpha^4\frac{Z}{6m_e^2}\;
    \left[
       \frac{2(\mathbf{s}_e\boldsymbol{\mu}_a)}{r_a^4}
       +\frac{r_a^2(\mathbf{s}_e\boldsymbol{\mu}_a)
             \!-\!3(\mathbf{r}_a\mathbf{s}_e)(\mathbf{r}_a\boldsymbol{\mu}_a)}
                                                             {r_a^6}
    \right],
\end{equation}
\end{subequations}
where $\boldsymbol{\mu}_I=\boldsymbol{\mu}_1+\boldsymbol{\mu}_2$.
It is convenient to separate the effective Hamiltonian into two terms:
scalar and tensor,
\begin{equation}
\begin{array}{@{}r@{\;}l}
H_s^{(6)} & \displaystyle
 = \alpha^4\frac{Z}{3m_e}\frac{\mu_p}{m_p}
   \left[
      -\frac{1}{4m_e^2}\Bigl\{p_e^2,4\pi
         \left[
            \delta(\mathbf{r}_1)\!+\!\delta(\mathbf{r}_1)
         \right]\Bigr\}
      +\frac{Z}{2m_e}
      \left(
         \frac{1}{r_1^4}+\frac{1}{r_2^4}
         +\frac{2\mathbf{r}_1\mathbf{r}_2}{r_1^3r_2^3}
      \right)
   \right]
   \left(\mathbf{s}_e\!\cdot\mathbf{I}\right)
\\[3mm]
H_t^{(6)} & \displaystyle
 = \alpha^4\frac{Z}{6m_e}
   \Biggl[
      \frac{r_1^2(\mathbf{s}_e\boldsymbol{\mu}_2)\!-\!
         3(\mathbf{r}_1\mathbf{s}_e)(\mathbf{r}_1\boldsymbol{\mu}_1)}{r_1^6}
      +\frac{r_2^2(\mathbf{s}_e\boldsymbol{\mu}_2)\!-\!
         3(\mathbf{r}_2\mathbf{s}_e)(\mathbf{r}_2\boldsymbol{\mu}_2)}{r_2^6}
\\[3mm] & \displaystyle \hspace{20mm}
      +\frac{(\mathbf{r}_1\mathbf{r}_2)(\mathbf{s}_e\boldsymbol{\mu}_I)
          \!-\!3(\mathbf{r}_1\mathbf{s}_e)(\mathbf{r}_2\boldsymbol{\mu}_2)
          \!-\!3(\mathbf{r}_2\mathbf{s}_e)(\mathbf{r}_1\boldsymbol{\mu}_2)}
                                                       {r_1^3r_2^3}
   \Biggr].
\end{array}
\end{equation}
$H_t^{(6)}$ has a finite expectation value, and since it does not contribute
to $b_F$, its consideration will be omitted in what follows. The divergent
terms are encountered only in the scalar Hamiltonian.

\subsection{Separating divergences in the second order contribution}

The operators which appear on the left and on the right of the second
order iteration are:
\[
H_B^{(1)}=4\pi\Bigl(\delta(\mathbf{r}_1)+\delta(\mathbf{r}_2)\Bigr),
\qquad
H_B^{(2)}=-\frac{\mathbf{p}_e^4}{8m_e^3}
          +\frac{Z}{8m_e^2}\,
             4\pi\Bigl(\delta(\mathbf{r}_1)+\delta(\mathbf{r}_2)\Bigr).
\]
Now we have to separate the singular part using the method outlined in the hydrogen case. We set :
\begin{equation}\label{Hplus_WFdelta}
\begin{array}{@{}l}
\displaystyle
(E_0-H_0)\Psi_B^{(1)}=
Q\left[4\pi\,\Bigl(\delta(\mathbf{r}_1)+\delta(\mathbf{r}_2)\Bigr)\right]\Psi_0
\\[3mm]\displaystyle
\Psi_B^{(1)} = 2m_e\left[-\frac{1}{r_1}-\frac{1}{r_2}\right]\Psi_0
               +\tilde{\Psi}_B^{(1)}
             = U_1\Psi_0+\tilde{\Psi}_B^{(1)},
\qquad
U_1 = \frac{2m_e}{Z}V,
\\[4mm]\displaystyle
H'^{(1)}_{B}=-(E_0-H_0)U_1-U_1(E_0-H_0)+H_B^{(1)}.
\end{array}
\end{equation}

Similarly, one gets for $H^{(2)}_{B}$:
\begin{equation}\label{Hplus_WFbreit}
\begin{array}{@{}l}
\displaystyle
(E_0-H_0)\Psi_B^{(2)}=
Q\left[
   -\frac{p^4}{8m_e^3}
   +\frac{Z\pi}{2m_e^2}\Bigl(\delta(\mathbf{r}_1)+\delta(\mathbf{r}_2)\Bigr)
\right]\Psi_0
\\[3mm]\displaystyle
\Psi_B^{(2)} = \frac{Z}{4m_e}
                  \left[-\frac{1}{r_1}-\frac{1}{r_2}\right]\Psi_0(r)
                               +\tilde{\Psi}_B^{(2)}
             = U_2\Psi_0+\tilde{\Psi}_B^{(2)},
\qquad
U_2 = -\frac{1}{4m_e}V.
\\[4mm]\displaystyle
H'^{(2)}_{B}=-(E_0-H_0)U_2-U_2(E_0-H_0)+H_B^{(2)}
\end{array}
\end{equation}
Applying these transformations to the second order iteration term we
arrive at
\begin{equation}\label{eq:2iteration_ion}
\begin{array}{@{}r@{\;}l}
\Delta E_A
& = \displaystyle
\alpha^4\frac{2}{3m_e}\frac{\mu_p}{M_p}
\Bigl(
   \left\langle
      \Psi_0\Big|\left(H_B^{(2)}
         -\left\langle H_B^{(2)}\right\rangle\right)U_1\Big|\Psi_0
   \right\rangle+
   \left\langle
      \Psi_0\Big|U_2({H'_{B\!}}^{(1)}
      -\langle{H'_{B\!}}^{(1)}\rangle)\Big|\Psi_0
   \right\rangle
\\[3mm] & \displaystyle \hspace{45mm}
   +\left\langle
      \Psi_0\Big|{H'_{B\!}}^{(2)}Q(E_0-H_0)^{-1}Q{H'_{B\!}}^{(1)}\Big|\Psi_0
   \right\rangle
\Bigr)
\left\langle \mathbf{s}_e\!\cdot\mathbf{I} \right\rangle.
\end{array}
\end{equation}
Again we pick out the first two terms which can be recast in the form of an effective
Hamiltonian:
\begin{equation}
\begin{array}{@{}r@{\;}l}
\displaystyle
H'^{(6)} & \displaystyle
 = \alpha^4\frac{1}{3Zm_e}\frac{\mu_p}{M_p}
   \biggl[
      -\frac{p^4V\!+\!Vp^4}{4m_e^2}-\frac{(Vp^2V)}{2m_e}
      -V^3+E_0\>V^2-4m_e\left\langle H_B^{(2)}\right\rangle V
      +\frac{Z\bigl\langle H_B^{(1)}\bigr\rangle V}{2m_e}
   \biggr]
   \left( \mathbf{s}_e\!\cdot\mathbf{I} \right)
\end{array}
\end{equation}
Its expectation value can be rewritten as follows:
\begin{equation}
\begin{array}{@{}r@{\;}l}
\displaystyle
\left\langle H'^{(6)}\right\rangle & \displaystyle
 = \alpha^4\frac{1}{3Zm_e}\frac{\mu_p}{M_p}
   \biggl[
      -\frac{2\pi Z\left\langle V
            \bigl[\delta(\mathbf{r}_1)\!+\!\delta(\mathbf{r}_2)\bigr]
            \right\rangle}{m_e}
      -\left\langle V^3\right\rangle
      +\frac{\left\langle \mathbf{p}V^2\mathbf{p} \right\rangle}{2m_e}
      +3E_0\left\langle V^2\right\rangle
      -2E_0^2\left\langle V\right\rangle
\\[1mm] & \hspace{65mm} \displaystyle
      -4m_e\left\langle H_B^{(2)}\right\rangle \langle V \rangle
      +\frac{Z\bigl\langle H_B^{(1)}\bigr\rangle \langle V \rangle}{2m_e}
   \biggr]
   \left\langle \mathbf{s}_e\!\cdot\mathbf{I} \right\rangle,
\end{array}
\end{equation}
and the divergent terms, the first two terms, are now written explicitly.

\subsection{Removing divergences and final expressions}

The remaining part of the second order iteration contribution (last term
in Eq.~(\ref{eq:2iteration_ion}) is now finite
\begin{equation}\label{H2plus_H4p_iter}
\Delta E'^{(6)}_A =
   \alpha^4\frac{2}{3m_e}\frac{\mu_p}{M_p}
   \left\langle
      \Psi_0\Big|{H'_{B\!}}^{(2)}Q(E_0-H_0)^{-1}Q{H'_{B\!}}^{(1)}\Big|\Psi_0
   \right\rangle
   \left\langle \mathbf{s}_e\!\cdot\mathbf{I} \right\rangle,
\end{equation}
where ${H'_{B\!}}^{(1)}$ and ${H'_{B\!}}^{(2)}$ are defined above.

Summing up $\left\langle H_S^{(6)}\right\rangle$ and
$\left\langle H'^{(6)}\right\rangle$ we get a finite expression as well
\begin{equation}\label{H2plus_H6p}
\begin{array}{@{}l}
\displaystyle
\Delta E'^{(6)}_B =
   \alpha^4\frac{1}{3Zm_e}\frac{\mu_p}{M_p}
   \biggl[
      \frac{\left\langle \mathbf{p}V^2\mathbf{p} \right\rangle}{m_e}
      +2E_0\left\langle V^2\right\rangle
      -2E_0^2\left\langle V\right\rangle
      -\frac{E_0}{m_e}\>4\pi Z
          \left\langle
             \delta(\mathbf{r}_1)\!+\!\delta(\mathbf{r}_2)
          \right\rangle
\\[3mm]\hspace{30mm}\displaystyle
      -4m_e\left\langle H_B^{(2)}\right\rangle \langle V \rangle
      +\frac{Z\bigl\langle H_B^{(1)}\bigr\rangle \langle V \rangle}{2m_e}
   \biggr]
   \left\langle \mathbf{s}_e\!\cdot\mathbf{I} \right\rangle.
\end{array}
\end{equation}

\begin{figure}[t]
\includegraphics*[width=0.4\textwidth]{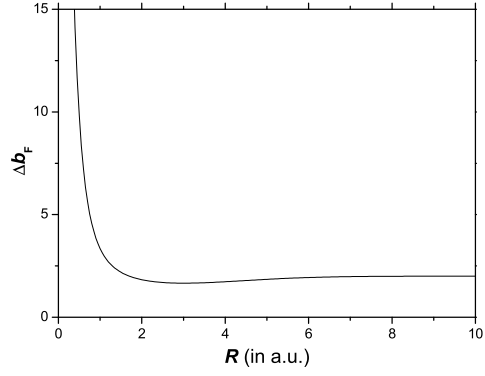}
\caption{Adiabatic effective potential for the relativistic correction of
order $m\alpha^6(m/M)$ to the spin-spin interaction coefficient $b_F$.
Energy scale for $\Delta b_F$ is in $(\alpha^4\,\mu_p/M_p)\times
(1\>\mbox{a.u.})$.} \label{fig:HFS_rel}
\end{figure}

\begin{table}[t]
\begin{center}
\begin{tabular}{@{\hspace{2mm}}c@{\hspace{5mm}}r@{\hspace{4mm}}r@{\hspace{6mm}}r@{\hspace{4mm}}r@{\hspace{2mm}}}
\hline\hline
\vrule width0pt height10pt depth4pt
 & \multicolumn{2}{c}{$L=1$} & \multicolumn{2}{c}{$L=3$} \\
\cline{2-5}
\vrule width0pt height10pt depth4pt
 & \cite{HFS06}~~~~ & new~~~ & \cite{HFS06}~~~~ & new~~~ \\
\hline
\vrule width0pt height10pt depth4pt
$v=0$ & 922.9918 & 922.9168 & 917.5911 & 917.5167 \\
$v=1$ & 898.8091 & 898.7371 & 893.7545 & 893.6831 \\
$v=2$ & 876.4542 & 876.3851 & 871.7277 & 871.6592 \\
$v=3$ & 855.8124 & 855.7460 & 851.3984 & 851.3325 \\
$v=4$ & 836.7835 & 836.7197 & 832.6682 & 832.6049 \\
\hline\hline
\end{tabular}
\end{center}
\caption{Results of numerical calculations for low ro-vibrational states
of the spin-spin interaction coefficient $b_F$ (in MHz).}
\label{tab:results}
\end{table}

\section{Results and conclusion}

\begin{table}[t]
\begin{center}
\begin{tabular}{@{\hspace{2mm}}c@{\hspace{5mm}}c@{\hspace{5mm}}c@{\hspace{5mm}}c@{\hspace{2mm}}}
\hline\hline
\vrule width0pt height10pt depth4pt
$v$ & \cite{HFS06} & this work & experiment \\
\hline
\vrule width0pt height10pt depth4pt
4   &  836.784   &  836.720  &  836.729  \\
5   &  819.280   &  819.219  &  819.227  \\
6   &  803.227   &  803.167  &  803.175  \\
7   &  788.558   &  788.501  &  788.508  \\
8   &  775.221   &  775.166  &  775.172  \\
\hline\hline
\end{tabular}
\end{center}
\caption{Comparison of the spin-spin interaction coefficient $b_F$ (in
MHz) with experiment. Ref.~\cite{HFS06} is the Breit-Pauli approximation
with account of the electron anomalous magnetic moment. $L=1$.}\label{comparison}
\end{table}

\begin{table}[t]
\begin{center}
\begin{tabular}{@{\hspace{2mm}}c@{\hspace{5mm}}r@{}l@{\hspace{5mm}}r@{}l@{\hspace{2mm}}}
\hline\hline
\vrule width0pt height10pt depth4pt
contribution & $v=4$~~& & $v=5$~~ & \\
\hline
\vrule width0pt height10pt depth4pt
$b_F$ \cite{HFS06}   &  836.7835 &    &  819.2801 &    \\
$(Z\alpha)^2$        &    0.0510 &    &    0.0511 &    \\
$\alpha(Z\alpha)$    & $-$0.0804 &    & $-$0.0787 &    \\
$\alpha(Z\alpha)^2\ln^2(Z\alpha)$
                     & $-$0.0067 &    & $-$0.0065 &    \\
$\Delta E_Z$         & $-$0.0335 &(5) & $-$0.0328 &(5) \\
$\Delta E^p_R$       &    0.0049 &(1) &    0.0045 &(1) \\
$\Delta E_{\rm pol}$ &    0.0012 &(5) &    0.0011 &(5) \\
\hline
\vrule width0pt height10pt depth4pt
$b_F$(new)           &  836.7197 &(10)&  819.2187 &(10)\\
\hline\hline
\end{tabular}
\end{center}
\caption{A summary of contributions to the spin-spin interaction
coefficient $b_F$ (in MHz).}\label{tab:contrib}
\end{table}

Results of numerical calculation as a function of a bond length for the
relativistic correction to the HFS within the framework of the two-center
problem are shown on Fig.~\ref{fig:HFS_rel}. In our study we use the
variational exponential expansion introduced in \cite{JCP06}. In fact, the
adiabatic effective potentials from Eq.~(\ref{H2plus_H6p}) have already
been obtained in the previous work \cite{KorJPB07} and only the second
order perturbation term (Eq.~(\ref{H2plus_H4p_iter})) with modified
operators $H'^{(1)}_B$ and $H'^{(2)}_B$ require some additional numerical
efforts.  The potential of the total effective Hamiltonian $\Delta
E'^{(6)}_B$ tends to zero when $R\to0$, or $R\to\infty$, as it may be
expected from the analysis of the hydrogen atom ground state HFS.

The relative numerical accuracy of the potential curve plotted on
Fig.~\ref{fig:HFS_rel} is estimated to be $\sim\!10^{-5}$, however the
adiabatic approximation itself limits the final uncertainty of the
relativistic contribution of the $m\alpha^6(m/M)$ order to the spin-spin
interaction coefficient $b_F$ to be about $0.1$ kHz (3-4 significant
digits in $\Delta b_F$). The other contributions which are described by
Eqs.~(\ref{eq:aZa})--(\ref{eq:pol}) may be obtained using the previously
calculated mean values of the delta function operators \cite{KorPRA06}.
The final results for the new theoretical value of the coefficient $b_F$
for the low ro-vibrational states are presented in Table
\ref{tab:results}.

An experimental value for $b_F$ can be uniquely calculated by using the
mixing parameters \cite{PRA08a} of the states $(F,J)$:
$(1/2,1/2)\leftrightarrow(3/2,1/2)$ and
$(1/2,3/2)\leftrightarrow(3/2,3/2)$, to restore the structure of pure
$F=1/2$ and $F=3/2$ multiplets and then take a difference between
statistically averaged splittings of these multiplets. In
Table~\ref{comparison} a comparison with experiment is given. As it may be
seen the newly obtained results improve the agreement with the experiment
by about a factor of 6. The error bars for $b_F$ from the experimental
data are to be about 1 kHz as it follows from the claimed accuracy of
Ref.~\cite{Jeff69}. On the other hand from the comparison with the hydrogen
atom case, the theoretical uncertainty should be no more than 2-3 kHz.
That indicates substantial discrepancy between theory and experiment of
about 6--9 kHz.

In order to try to explain this discrepancy we have checked several
effects which may have impact on the spin-spin interaction. The leading
order retardation effects in the nonrelativistic interaction region
\cite{Pac98} as well as the cross terms of the second order perturbation
(when electron interacts with both protons in $\mbox{H}_2^+$) for the
proton structure dependent contributions are estimated either equal to
zero or negligibly small. We have also analyzed the effect of the $g/u$
symmetry breaking which is essential for high $v$ states, say, for $v=19$,
it leads to a few MHz shift in energy \cite{Moss}. However for the states
below $v=10$ this effect is smaller than 1 kHz and, thus, the gap between
theory and experiment can not be accounted for by the $g/u$ mixing. A
possible explanation is that the higher order corrections
($\alpha^2(m/M)E_F$ and $\alpha^3E_F$) may give significant contribution.

In conclusion, the consideration of $m \alpha^6$ and (partially) $m \alpha^7$
order corrections, as well as proton finite size effects, has allowed to
improve significantly the agreement with experiment, to about 10 ppm. The
remaining discrepancy is somewhat larger than expected from comparison
with the hydrogen atom, and further theoretical work to improve the HFS
intervals is needed. In any case, a new independent experiment is highly
desirable.

\section{Acknowledgments}

This work was supported by l'Universit\'e d'Evry Val d'Essonne and by
R\'egion Ile-de- France. V.I.K. acknowledges support of the Russian
Foundation for Basic Research under Grant No. 08-02-00341. Laboratoire
Kastler Brossel de l'Universit\'e Pierre et Marie Curie et de l'Ecole
Normale Sup\'erieure is UMR 8552 du CNRS. We wish to thank K.~Pachucki for
helpful comments and discussion.

\end{document}